\documentclass[showpacs, twocolumn]{revtex4-1}
\usepackage{graphicx}
\usepackage{amsmath}
\usepackage{appendix}

\begin{document}

\title{Moving Bose mixtures with dipole-dipole interactions}

\author{Bakhta Ch\'erifi$^{1,2}$ and Abdel\^{a}ali Boudjem\^{a}a$^{1,2}$}
\affiliation{
$^1$ Department of Physics, Faculty of Exact Sciences and Informatics, Hassiba Benbouali University of Chlef, P.O. Box 78, 02000, Ouled-Fares, Chlef, Algeria.\\
$^2$Laboratory of Mechanics and Energy, Hassiba Benbouali University of Chlef, P.O. Box 78, 02000, Ouled-Fares, Chlef, Algeria.}
\email {a.boudjemaa@univ-chlef.dz}


\begin{abstract}
We study the properties of moving  uniform dipolar Bose-Bose mixtures using the full Hartree-Fock-Bogoliubov theory. 
The analytical and numerical calculations emphasize that the competition between the relative motion of two fluids and the interspecies dipole-dipole interactions
may affect the behavior of the condensed depletion, the anomalous density, the ground-state energy and second-order pair correlation.
It is found that in the lower branch of the mixture, these quantities are unimportant and present an unconventional behavior.
\end{abstract}


\maketitle

\section{Introduction}

Ultracold Bose mixtures have triggered enormous interest both experimentally \cite{Mya,Hall,Mad,Pap,Sug,Mog, McC,Ler,Pasq,Wack,Wang,Igor} and theoretically  
\cite{Ho,Esry,Timm,Ao,Sinatra,Koz,Band, Jez, Svid,Kasa,Roy,Lee,Boudj,Boudj1, BoudjGSK, BoudjA, Ota} over the last two decades.
The properties of homogeneous binary Bose-Einstein condensates  (BECs)  were analyzed in \cite {Larsen, Bass, YNep, Sor, Tom} using the Bogoliubov theory. 
These quantum ensembles open exciting possibilities for the investigation of intriguing interplay between the two condensates  
(see e.g. \cite{ Esry, Timm, Jez, Svid,Lee} and references therein), 
and beyond mean-field effects i.e. quantum droplets  (see e.g. \cite{Petrov, Cab,Sem, Err,Boudj2,Boudj3} and references therein).

Recent advances in experimental techniques have paved the way for condensates with dipole-dipole interactions (DDI) 
that is both long ranged and anisotropic and it can be also attractive and repulsive \cite{Pfau,ming,erbium,lu}.
Dipolar atomic systems  provide a versatile platform to study quantum many-body effects (see for review \cite{Pfau,Carr,Baranov,Pupillo2012}).
Furthermore, dipolar Bose-Bose mixtures are ideally suited to the study of many extraordinary quantum phenomena such as 
rotonization and the miscible-immiscible transition \cite{Wilson,Kumar}, solitons \cite{Adhik1},  exotic supersolidity (see e.g. \cite{Wilson1, Yong}),  superfluidity \cite{Boudj4},
low-temperature properties  \cite{Past, Boudj5}. Recently realized mixtures of  Er-Dy atoms \cite{Ilz} open new avenues for the
exploration of quantum self-bound droplets  \cite{Boudj5,Biss}.

Motivated by the above interesting works, we present in this paper a comprehensive discussion of the ground-state properties of moving dipolar Bose-Bose mixtures.
Such quantum mixtures may afford a fascinating glimpse into the ultracold gases due to the competition between the interspecies dipolar interactions and the relative motion. 
They may help us to understand the quantum transport and to uncover new physical effects in dipolar mixtures.
In the nondipolar case, Bose mixtures  exhibit a complex motion that tends to preserve the total density 
but quickly damped to a stationary state with non-negligible component overlap.
They have also the possibility to collide and move through each other \cite{Hall, Koz, Mad, Band}.
The presence of the relative motion of two species may lead to strongly influence the stability and the quantum fluctuations in both
clean \cite{Yuk} and dirty mixtures \cite{Boudj6}.

In this paper we introduce a generalized Hartree-Fock-Bogoliubov (HFB) theory of the weakly interacting dipolar Bose mixtures of moving components.
The approach we develop here is conserving and gapless enables the self-consistent treatment of the Bogoliubov excitations energy, 
the condensate depletion, the anomalous correlation (pairing), and the ground-state energy  for both single and binary BECs.
Additionally, we accurately examine the role played by the interplay of DDI and the velocity of two BECs on the second-order correlation function.

The profiles of the different physical quantities are computed numerically for parameters relevant to the recent experiments.
In the single BEC, the relative motion  gives rise to corrections that decrease the depletion and enhance the ground-state energy.
On the other hand, our results reveal that the interplay of the interspecies dipolar interactions and the relative motion of two fluids 
may lower the condensed depletion and  increase the anomalous density and the ground-state energy in each species. 
In the lower branch which corresponds to the spin excitations (hard modes), the depletion, the anomalous density and the energy exhibit an unusual behavior.
Such an energy growth can be explained by the fact that moving dipolar Bose fluids gain an additional energy $\propto v^2$, where $v$ is the velocity of system.
It is found that the relative motion of two BECs, the interspecies dipolar interactions, and the dipoles orientation may also affect the second-order correlation function and thus, 
the coherence of the mixture.

The rest of the paper is organized as follows. In section \ref {Model}, we introduce the full HFB formalism for moving dipolar binary BECs.
In section \ref{MSDB} we discuss the properties of a moving single dipolar BEC. 
In section \ref{MDBM} we extend our study to a moving dipolar Bose mixture, where the behavior of the quantities of interest are computed for 
parameters relevant to recent experiment.
Section \ref{PCF} is devoted to the analysis of the second-order correlation function of the mixture under consideration.
Our conclusions are drawn in section \ref{Conc}.

\section {Formalism} \label{Model}

\subsection{Generalized coupled Gross-Pitaevskii equations}

We consider weakly interacting  two-component dipolar BECs with equal masses $m_1=m_2=m$ confined in an external trap $U({\bf r})$.
The generalized nonlocal coupled GP equations for Bose-Bose mixtures read \cite{Boudj5}
\begin{widetext}
\begin{align}\label{GPE}   
i\hbar \dot{\Phi}_j ({\bf r},t)& =  \bigg [\frac{{\bf p}^2}{2m}+U({\bf r})\bigg] \Phi_j ({\bf r},t) +\int d{\bf r'} V_j({\bf r}-{\bf r'}) \bigg [ n_j ({\bf r'},t) \Phi_j({\bf r},t) \\
&+\tilde n_j ({\bf r},{\bf r'},t)\Phi_j ({\bf r'},t)  +\tilde m_j ({\bf r},{\bf r'},t)\phi_j^*({\bf r'},t) \bigg ] +\int d{\bf r'} V_{12} ({\bf r}-{\bf r'}) n_{3-j} ({\bf r'}) \Phi_j ({\bf r},t), \nonumber
\end{align}
\end{widetext}
where ${\bf p}= -i \hbar {\bf \nabla}$ is the kinetic-energy operator, $V_j ({\bf r})$ and $V_{12}({\bf r})$ are, respectively the intraspecies and interspecies two-body interactions.
The quantities $n_{cj}({\bf r})=|\Phi_j({\bf r})|^2$, $\tilde n_j ({\bf r})= \langle \hat {\bar\psi}_j^\dagger ({\bf r}) \hat {\bar\psi}_j ({\bf r}) \rangle $ and 
$\tilde m_j ({\bf r})= \langle \hat {\bar\psi}_j ({\bf r}) \hat {\bar\psi}_j ({\bf r}) \rangle $ are, respectively  the condensed, noncondensed and anomalous densities,
where $\hat{\bar \psi}_j({\bf r})=\hat\psi_j({\bf r})- \Phi_j({\bf r})$ is the noncondensed part of the field operator with $\Phi_j({\bf r})=\langle\hat\psi_j({\bf r})\rangle$ 
being the condensate wave-function. 
The total density in each component is given by $n_j({\bf r})=n_{cj}({\bf r})+\tilde n_j ({\bf r})$.
The terms $\tilde n_j ({\bf r, r'})$ and $\tilde m_j ({\bf r, r'})$ stand for the normal and the anomalous one-body density matrices
which represent the dipole exchange interaction between the condensate and noncondensate. 

The intraspecies two-body interaction potential is defined as:
\begin{equation}  \label{IntraPot}
V_j ({\bf r})=g_j\delta({\bf r})+d_j^2 \frac{1-3\cos^2\theta} { r^3},
\end{equation}
where $g_j=4\pi \hbar^2 a_j/m$ with $a_j$ being the intraspecies  $s$-wave scattering lengths.
The last term in Eq.(\ref{IntraPot}) accounts for the DDI potential where $d_j$ stands for the magnitude of the dipole moment of component $j$ and 
$\theta$ is the angle between the polarization axis and the relative separation of the two dipoles, it is supposed to be the same for both components.
The intraspecies dipole-dipole distance is defined as $r_{*j}= m d_j^2/\hbar^2$. \\
The interspecies two-body interactions potential reads
\begin{equation}  \label{InterPot}
V_{12}({\bf r})=g_{12}\delta({\bf r})+d_1 d_2 \frac{1-3\cos^2\theta} { r^3},
\end{equation}
where $g_{12}=g_{21}= 4\pi \hbar^2 a_{12}/m$ corresponds to the interspecies 
short-range part of the interaction, which is characterized by the interspecies  $a_{12}=a_{21}$ $s$-wave scattering lengths.
The interspecies dipole-dipole distance is $r_{*12}=r_{*21}= d_1 d_2 m/\hbar^2$.

For $r_{*1}=r_{*2}=r_{*12}=0$, Eqs.(\ref{GPE}) reduce to the coupled GP equations for finite-temperature nondipolar mixtures \cite{Boudj1}.
If $\tilde n_j=\tilde m_j=0$, one can reproduce the nonlcoal GP equations for binary condensates at zero temperature.

Now we assume a mixture of moving components, where each component moves with the same velocity $v_1 =v_2 =v$.
In such a case the Andreev-Bashkin effect (i.e. the existence of a non-zero entrainment between two species) \cite{AB} totally disappears.
The motion of the mixture is done by means of the Galilean transformation:
\begin{equation}\label{GP1}
	\phi({\bf r},t)=\phi({\bf r},t)\exp{ \left(\frac{i}{\hbar} m {\bf v} \cdot{\bf r}\right)}.
\end{equation}
According to Tisza and Landau \cite{Tisza, Land}, the Galilean invariance guarantees the superfluidity of the system.
Introducing Eq.(\ref{GP1}) into the set (\ref{GPE}), the above coupled GP equations can be rewritten as follows:
\begin{widetext}
\begin{align}\label{GPE1}   
i\hbar \dot{\Phi}_j ({\bf r},t) &=  \bigg [\frac{{\bf p}^2}{2m}+{\bf p} \cdot{\bf v}+\frac{1}{ 2}m v^2+U({\bf r})\bigg] \Phi_j ({\bf r},t) +\int d{\bf r'} V_{12} ({\bf r}-{\bf r'}) n_{3-j} ({\bf r'}) 
\Phi_j ({\bf r},t)\\
&+\int d{\bf r'} V_j({\bf r}-{\bf r'}) \bigg [ n_j ({\bf r'},t) \Phi_j({\bf r},t) +\tilde n_j ({\bf r},{\bf r'},t)\Phi_j ({\bf r'},t)  +\tilde m_j ({\bf r},{\bf r'},t)\phi_j^*({\bf r'},t) \bigg ].  \nonumber
\end{align}
\end{widetext}
For $\tilde n_j=\tilde m_j=0$, one recovers the coupled GP equations for moving dual BECs at zero temperature \cite{Yuk,Boudj6}.

\subsection{Collective excitations}

Upon linearizing Eq.(\ref{GPE1})  around a static solution $\Phi_0$, utilizing the transformation
$\Phi_j({\bf r},t)=\left[\Phi_{0j} ({\bf r})+\delta\Phi_j({\bf r},t) \right]\exp{\left(- i\mu_j t/\hbar \right)}$,
where $\mu_j$ are chemical potentials related with bosonic components,
and  $\delta\phi_j({\bf r},t)=u_{j \bf p}({\bf r})\exp({-i\varepsilon_{\bf p}  t/\hbar}) +v_{j \bf p}({\bf r})\exp({i\varepsilon_{\bf p} t/\hbar})$ are small quantum fluctuations
with $\varepsilon_{\bf p}$ being the Bogoliubov excitations energy.
The quasi-particle amplitudes $ u_{j \bf p}({\bf r})$ and $v_{j\bf p}({\bf r})$ satisfy the generalized nonlocal Bogoliubov-de-Gennes (BdG) equations:
\begin{widetext}
\begin{align}
\varepsilon_{\bf p}  u_{j {\bf p} } ({\bf r}) &= \hat {\cal L}_j u_{j{\bf p} } ({\bf r})+ \int d{\bf r'} V({\bf r}-{\bf r'}) n_j ({\bf r},{\bf r'}) u_{j{\bf p} } ({\bf r'}) 
+ \int d {\bf r'}  V({\bf r}-{\bf r'}) \bar m_j  ({\bf r},{\bf r'}) v_{j{\bf p} } ({\bf r'})  \label{BdG1} \\
&+  \int d {\bf r'}  \Phi_{0,3-j}({\bf r'}) V_{12}({\bf r}-{\bf r'}) \Phi_{0j}({\bf r}) u_{3-j,{\bf p} } ({\bf r'}) + \int d {\bf r'}  \Phi_{0,3-j}^*({\bf r'}) V_{12}({\bf r}-{\bf r'}) 
\Phi_{0j}({\bf r}) v_{3-j,{\bf p} } ({\bf r'}),\nonumber \\ 
-\varepsilon_{\bf p}  v_{j {\bf p}} ({\bf r}) &= \hat {\cal L} v_{j {\bf p}} ({\bf r})+ \int d{\bf r'} V({\bf r}-{\bf r'}) n_j ({\bf r},{\bf r'}) v_{j{\bf p}} ({\bf r'}) 
+ \int d {\bf r'}  V({\bf r}-{\bf r'}) \bar m_j  ({\bf r},{\bf r'})  u_{j{\bf p}} ({\bf r'}), \label{BdG2}\\
&+  \int d {\bf r'}  \Phi_{0,3-j}({\bf r'}) V_{12}({\bf r}-{\bf r'}) \Phi_{0j}({\bf r}) v_{3-j,{\bf p}} ({\bf r'})+  \int d {\bf r'}  \Phi_{0,3-j}^*({\bf r'}) V_{12}({\bf r}-{\bf r'}) \Phi_{0j}({\bf r}) 
u_{3-j,{\bf p}} ({\bf r'}),\nonumber 
\end{align}
\end{widetext}
where 
$\hat {\cal L}_p={\bf p}^2/2m+{\bf p}\cdot{\bf v}+m v^2/2+U({\bf r})+ \int d{\bf r'} V({\bf r}-{\bf r'}) n_j ({\bf r'})+\int d{\bf r'} V_{12}({\bf r}-{\bf r'}) n_{3-j} ({\bf r'})-\mu$, 
$n_j ({\bf r},{\bf r'})= \Phi_{0j}^*({\bf r'}) \Phi_{0j}({\bf r})+ \tilde n_j ({\bf r},{\bf r'})$ and $\bar m_j  ({\bf r},{\bf r'})= \Phi_{0j}({\bf r'}) \Phi_{0j}({\bf r}) +\tilde m_j  ({\bf r},{\bf r'})$. 
The nonlocal BdG Eqs.(\ref{BdG1}) and (\ref{BdG2}) describe the collective excitations of moving dipolar Bose mixtures.

Now let us consider the case of a uniform mixture $U({\bf r})=0$.
In such a case the wavefunctions are real-valued  ($\Phi_{0j}=\Phi_{0j}^*=\sqrt{n_{cj}}$, and $\Phi_{0\,3-j}=\Phi_{0\,3-j}^*=\sqrt{ n_{c3-j}}$), and 
$\delta\phi_j({\bf r},t)=u_{j {\bf p}}\exp({i{\bf p \cdot r}/\hbar-i\varepsilon_{\bf p} t/\hbar}) +v_{j{\bf p}}\exp({i{\bf p \cdot r}/\hbar+i\varepsilon_{\bf p} t/\hbar})$.
The Fourier transforms of interaction potentials (\ref {IntraPot})  and (\ref {InterPot}) are written as:
$\tilde V_j(\mathbf p)=g_j [1+\epsilon_j^{dd} (3\cos^2\theta-1)]$, and $\tilde V_{12}(\mathbf p)=g_{12} [1+\epsilon_{12}^{dd} (3\cos^2\theta-1)]$,
where $\epsilon_j^{dd}=r_{*j}/3a_j$ and $\epsilon_{12}^{dd}=r_{*12}/3a_{12}$ and $\theta$ is the angle between the vector $\mathbf p$ and the polarization direction.

The chemical potential in each species is given according to Eq.(\ref{GPE1}) by
\begin{align} \label{chim0}
\mu_j&=\frac{1}{2}m v^2+\tilde V_j(\mathbf p=0) n_j  +\tilde V_{12}(\mathbf p=0) n_{3-j} \\
&+\tilde V_j(\mathbf p) \big(\tilde n_j +\tilde m_j \big).  \nonumber
\end{align}
If one substitutes Eq.(\ref{chim0}) and the above Bogoliubov transformation into Eqs.(\ref{BdG1}) and (\ref{BdG2}),  the coupled BdG equations
in principle do not guarantee to give the best excitation frequencies due to the inclusion of the anomalous average which leads to the appearance of a gapped excitation spectrum.
One way to cure this problem is to use the condition $\tilde{m}/n_c \ll 1$ \cite{Boudj1,Boudj5}, which is valid at low temperature 
and necessary to ensure the diluteness of the system. 
Doing so, the BdG equations take the form:
\begin{align}
	\varepsilon_{\bf p} u_{j{\bf p}}&=\left [E_{\bf p}+ {\bf p} \cdot {\bf v}+\tilde V_j(\mathbf p) n_{cj}\right] u_{j{\bf p}}+\tilde V_j(\mathbf p) n_{cj}v_{j{\bf p}} \label{BdG11}  \\
&+\tilde V_{12}(\mathbf p)\sqrt{n_{cj} n_{c3-j}}(u_{j{\bf p}}+v_{j{\bf p}}),  \nonumber
\end{align}
and 
\begin{align} 
	-\varepsilon_{\bf p} v_{j{\bf p}}&=\left[E_{\bf p}+{\bf p} \cdot {\bf v}+\tilde V_j(\mathbf p) n_{cj}\right] v_{j{\bf p}}+ \tilde V_j(\mathbf {\bf p}) n_{cj} u_{j{\bf p}}  \\\label{BdG22} 
&+\tilde V_{12}(\mathbf p)\sqrt{n_{cj} n_{c3-j}}(u_{j{\bf p}}+v_{j{\bf p}}), \nonumber
\end{align}
where $ E_{\bf p}= {\bf p}^2/ 2m$ is the free particle energy.

From now on we assume a symmetric mixture with $n_1=n_2=n$, $n_{c1}=n_{c2}=n_c$ and $\tilde V_1(\mathbf p)=\tilde V_2(\mathbf p)=\tilde V(\mathbf p)$.

The quasiparticle amplitudes $u_{{\bf p}\pm}, v_{{\bf p}\pm}$ can be written as:
\begin{equation} \label{BAmp}
	u_{{\bf p}\pm},v_{{\bf p}\pm} =\frac{ 1}{ 2} \left (\sqrt{ \frac{\varepsilon_{{\bf p}\pm} } {E'_{\bf p}} }\pm\sqrt{ \frac{E'_{\bf p}} {\varepsilon_{{\bf p}\pm}} } \right),
\end{equation}
where $E'_{\bf p}=E_{\bf p}+p v \cos\alpha$ with $\alpha$ being the angle between the two vectors ${\bf p}$ and  ${\bf v}$. 
The Bogoliubov excitations spectrum reads
\begin{equation} \label{Bog}
	\varepsilon_{{\bf p}\pm}=\sqrt{E_{\bf p}^{'2}+2 E'_{\bf p}  n_c \delta  V_{\pm}(\mathbf p) },
\end{equation}
where $\delta  V_{\pm}(\mathbf p)=\tilde V(\mathbf p) [1\pm \tilde V_{12}(\mathbf p)/\tilde V(\mathbf p)]$.
Evidently, the sepctrum (\ref{Bog}) is composed of two branches namely:
the upper branch $\varepsilon_{{\bf p}+}$, known as the hard mode, corresponds to the spin excitations  and lower energy branch $\varepsilon_{{\bf p}-}$ 
is known as the soft branch and corresponds to the density excitations.
For $v=0$, one reproduces the standard density and spin excitations for immovable dipolar Bose mixtures: 
$\varepsilon_{{\bf p} \pm}={\bf p}\sqrt{({\bf p}/2m)^2+c_{s\pm}(\theta)}$ \cite{Boudj5},
where $c_{s+} (\theta)= \sqrt{n_c\delta V_{+} (\mathbf p=0)/m}$, and $c_{s-}(\theta)= \sqrt{n_c\delta V_{-} (\mathbf p=0)/m}$ 
are the angular-dependent sound velocities in the density and spin channels, respectively (i.e. they acquire a dependence on the propagation
direction $\theta$, owing to the anisotropy of the DDI). For $\epsilon_j^{dd}=\epsilon_{12}^{dd}=0$, the spectrum of nondipolar Bose mixtures is recovered \cite{Boudj6}.

The stability of dipolar symmetric mixtures requires the conditions: $\epsilon_j^{dd} \leq 1$, and $\tilde V_{12}(\mathbf p) \leq \tilde V(\mathbf p)$.
For $\theta=\pi/2$, a stable mixture demands the inequality $(g_{12}/g) (1-\epsilon_{12}^{dd}) \leq (1-\epsilon_j^{dd})$.
Further necessary condition for the spectrum (\ref{Bog}) of such a moving structure possess positive solutions is $\alpha \in [0,2\pi]$.
Note that the stability condition could also be modified as a result of the relative motion of two liquids \cite{Yuk}, the temperature \cite{Boudj1,Ota} and disorder effects \cite{BoudjA}.

We assume henceforth that $\delta  V_{\pm}(\mathbf p)>0$ and $\alpha \in [0,2\pi]$. 
In this case, the minimum value possible for $v$ which must be kept positive,
can be obtained in lowest value of $p$: 
\begin{equation} \label{CritV}
v_c= c_{s\pm} (\theta),
\end{equation}
below which there is no solution for $v$. 
For $\epsilon_j^{dd}=\epsilon_{12}^{dd}=0$, this condition reduces to that obtained for a nondipolar Bose mixture \cite{Boudj6}.

\subsection{Fluctuations and equation of state}

At zero temperature, the condensate depletion and the anomalous density
for each component are defined, respectively as \cite{Boudj5}: $\tilde{n}_{\pm}=V^{-1} \sum_{\bf p} v_{{\bf p} \pm}^2$,
and $\tilde{m}_{\pm}=-V^{-1} \sum_{\bf p} u_{{\bf p} \pm} v_{{\bf p} \pm} $. Working in the thermodynamic limit, the sum over $p$ can be replaced  by the integral 
$\sum_{\bf p} \rightarrow V\int_{ 0}^{ \infty}  d{\bf p}/(2\pi\hbar)^3 $.
Thus, the noncondensed and anomalous densities take the form :
\begin{align}\label{Dep} 
\tilde n_{\pm}=\frac{1}{2}\int \frac{d {\bf p}}{ (2\pi \hbar)^3} \left[\frac{E'_{\bf p}+ mc_{s\pm}^2(\theta)} {\varepsilon_{{\bf p} \pm}}-1\right],
\end{align}
and
\begin{equation}\label{mDep} 
\tilde m_{\pm}=-\frac{1}{2}\int \frac{d {\bf p}}{ (2\pi \hbar)^3} \frac{mc_{s\pm}^2(\theta) } {\varepsilon_{{\bf p} \pm}}.
\end{equation}
The validity of the present approach requires the inequality: $\tilde n=\sum_{\pm} n_{\pm} \ll n$. 

Substituting $\tilde n_{\pm}$ and  $\tilde m_{\pm}$ from Eqs.(\ref{Dep}) and (\ref{mDep}) into Eq.(\ref{chim0}) we obtain for the chemical potential:
\begin{align} \label{chim01}
\mu_j&=\frac{1}{2}m v^2+\tilde V_j(\mathbf p=0) n_j  +\tilde V_{12}(\mathbf p=0) n_{3-j} \\
&+\frac{1}{2}\int \frac{d {\bf p}}{ (2\pi \hbar)^3} \tilde V_j(\mathbf p) \left[\frac{E'_{\bf p}} {\varepsilon_{{\bf p} \pm}}-1\right]. \nonumber
\end{align}
The leading term originates from the motion of two fluids, the second and third terms represent the mean-field contribution.
The last term accounts for the corrections to the chemical potential due to the Lee-Huang-Yang (LHY) quantum fluctuations stemming from the 
noncondesed and anomalous densities.

Expression (\ref{chim01}) allows one to calculate the ground-state energy via $E_j=\int dn_j \mu_j$. It can be explicitly written  as :
\begin{align} \label{Energy}
E_j&=\frac{1}{2}m v ^2 n_j+ \frac{1}{2}\tilde V_j(\mathbf p=0) n_j ^2 +\frac{1}{2}\tilde V_{12}(\mathbf p=0) n_{3-j}^2 \nonumber\\
&+E_{\text{LHY} \pm},
\end{align}
where 
\begin{equation}\label{LHYEgy}
E_{\text{LHY} \pm}=\frac{1}{2} \int \frac{d {\bf p}}{ (2\pi \hbar)^3} \big[\varepsilon_{ {\bf p} \pm} -E_{\bf p}'- mc_{s\pm}^2(\theta_{\bf p})\big],
\end{equation}
accounts for corrections to the ground-state energy owing to the LHY quantum fluctuations for Bose mixtures.\\
Equation (\ref{Energy}) shows that the energy of each component of a moving Bose mixture gains an additional term $m v^2 n_j/2$ 
in comparison with the energy of the system at rest. This is indeed in perfect consistent with the Tisza-Landau  theory of superfluidity \cite{Tisza, Land}.

Remarkably, the mean-field contribution to chemical potential (\ref{chim01}) and ground-state energy (\ref{Energy}) (second and third terms) 
acquire a dependence on the propagation direction owing to the anisotropy of the dipolar interaction. 
However, contributions arising from the LHY quantum fluctuations are isotropic as we will see in next sections.

One problem that commonly  confronted while implementing chemical potential (\ref{chim01}) and ground-state energy (\ref{Energy}) is the ultraviolet divergence 
originate from the contact potential which is valid only at low-momenta.  
To cure these difficulties one should use either the renormalization of the coupling constant \cite {Beleav,Griffin, peth,Boudj7} 
or  the dimensional regularization \cite{Boudj8, Anders,Yuk1}.

\section{Moving single dipolar BEC} \label {MSDB}

Let us start with a moving single  dipolar BEC i.e $\epsilon_{12}^{dd}=0$. In such a case, the excitations spectrum becomes 
$\varepsilon_{\bf p}=\sqrt{E_{\bf p}^{'2}+2 E'_{\bf p}  n_c g [1+\epsilon^{dd} (3\cos^2\theta-1)] }$, and the corresponding sound velocity takes the form
$c_s(\theta)= \sqrt{n_c V (\mathbf p=0)/m}$.

\begin{figure}[htb] 
\includegraphics[scale=0.7]{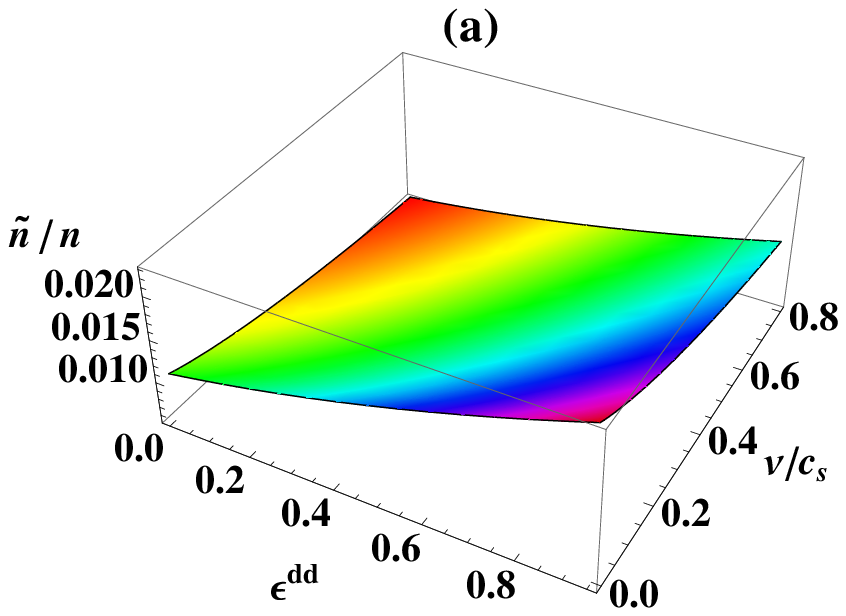}
\includegraphics[scale=0.7]{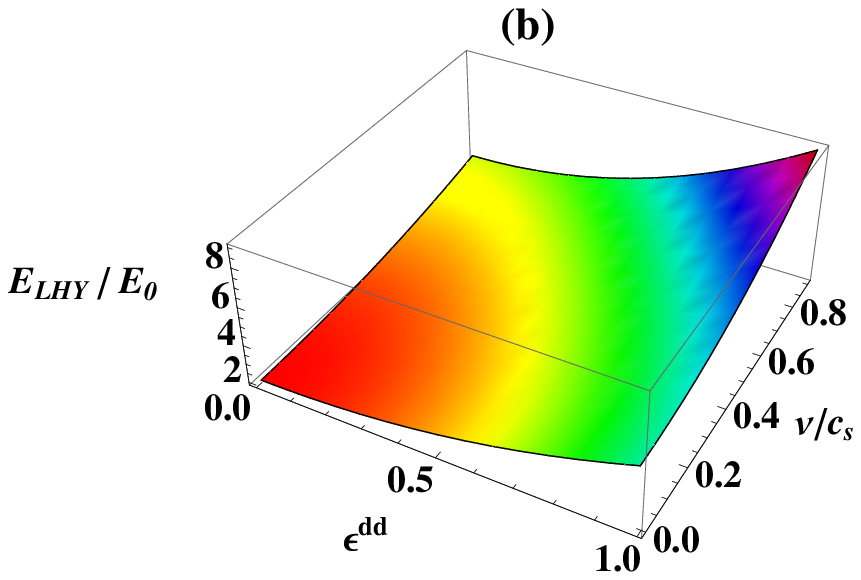}
\caption { (Color online) (a) Noncondensed density associated with the upper branch as a function of $\epsilon^{dd}$ and $v/c_s$.
(b) The LHY corrected energy as a function of $\epsilon_{12}^{dd}$ and $v/c_s$.
Parameters are: $a=141 a_0$ \cite{Tang}, $r_*=131 a_0$ \cite{Pfau1},  and $ n \approx 10^{20} m^{-3}$.}
\label{SB}
\end{figure}

To illustrate our generalized HFB formalism, we consider Dy BEC  with parameters: $a=141 a_0$ \cite{Tang},  $r_*=131 a_0$ \cite{Pfau1},  
$ n \approx 10^{20} m^{-3}$. We then solve numerically integrals (\ref{Dep}) and (\ref{LHYEgy}) for the depletion and the energy setting $\epsilon_{12}^{dd}=0$.
The results are shown in Fig.\ref{SB}.

Fig.\ref{SB}.(a) depicts that the condensate depletion $\tilde n$ increases with DDI while it decreases with the velocity of fluid.
The reason of such a decay can be understood from the fact that the noncondensed atoms hider the motion of atoms through the system.

The behavior of the LHY corrected-energy is captured in Fig.\ref{SB}.(b). 
We observe that $E_{\text{LHY}} $ increases with both the relative motion effects and the dipolar interactions. \\
For small velocities $v \ll v_c= c_{s\pm} (\theta)$, the energy can be easily computed by integrating term by term the power series
representation of integral (\ref{LHYEgy}).
To overcome the ultraviolet divergence, we use the dimensional regularization which is valid for weak interacions \cite{Boudj8, Anders,Yuk}.
We then obtain up to second-order in $v$
\begin{equation} \label{Energy2} 
E_{\text{LHY}} \simeq E_0 {\cal Q}_5(\epsilon^{dd}) +\frac{1}{2} m^*v^2,
\end{equation}
where $E_0=8V m^4 c_s^5/15 \pi^2\hbar^3$ is the famous LHY equation of state of an immovable condensate \cite{LHY},
and $m^*= 15 E_0 {\cal Q}_5(\epsilon^{dd})/(8 c_s^2)$.
The contribution of the DDI is expressed by the function ${\cal Q}_5(\epsilon^{dd})$, which is a special case $j=5$ of
 ${\cal Q}_j(\epsilon^{dd})=(1-\epsilon^{dd})^{j/2} {}_2\!F_1\left(-\frac{j}{2},\frac{1}{2};\frac{3}{2};\frac{3\epsilon^{dd}}{\epsilon^{dd}-1}\right)$, where ${}_2\!F_1$ 
is the hypergeometric function \cite{Boudj7,lime}. 
Note that functions ${\cal Q}_j(\epsilon^{dd})$ reach their maximal values for $\epsilon^{dd} \approx 1$ and become imaginary for $\epsilon_{dd}>1$.
Evidently, for a condensate with a pure contact interaction (${\cal Q}_5(\epsilon^{dd}=0)=1$) and for $v = 0$,  the energy (\ref{Energy2}) returns in to $E_0$.
For $v=0$, one recovers the energy of an immovable dipolar BEC \cite{Boudj7,lime}.
Equation (\ref{Energy2}) obviously shows that $E_{\text{LHY}}$ grows with both $v/c_s$ and DDI in agreement with our numerical findings shown in Fig.\ref{SB}.b.

As anticipated above, corrections due to the LHY quantum fluctuations and to the relative motion of fluid are isotropic (do not possess any dependence on the momentum direction) 
since they contains an integral over all the ${\bf p}$ modes.

\section{Moving dipolar Bose mixtures} \label {MDBM}

To gain deeper insights into the effects of the peculiar interplay of the relative motion and DDI, we consider  a Dy-Dy mixture  with parameters : $a=141 a_0$ \cite{Tang}, $a_{12}=115.1 a_0$, $r_*=131 a_0$ \cite{Pfau1},  $ n \approx 10^{20} m^{-3}$, and the relative interspecies dipolar interaction strength $\epsilon_{12}^{dd}$ can be adjusted by means of a Feshbach resonance.
It is worth stressing that our theory can be applied to all kinds of Bose mixtures. We then solve numerically the full integrals (\ref{Dep}) and (\ref{LHYEgy}).

\begin{figure}[htb] 
\includegraphics[scale=0.45]{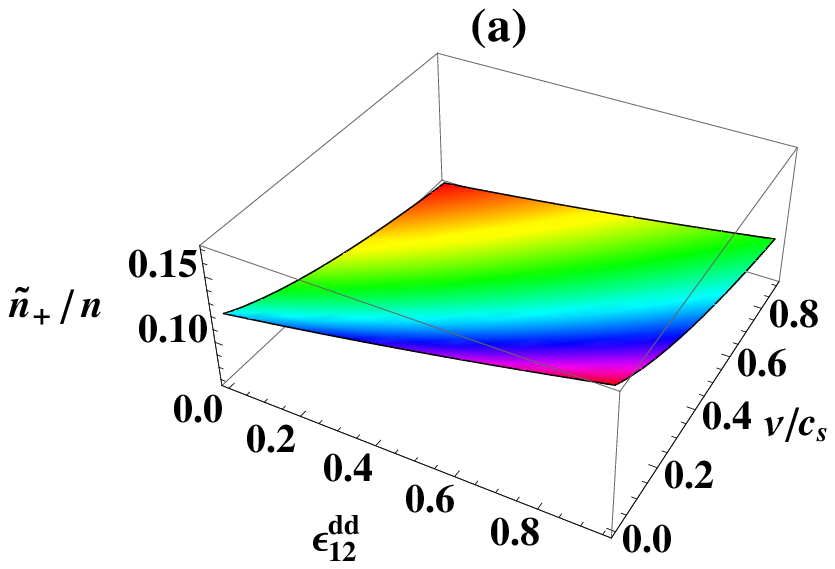}
\includegraphics[scale=0.45]{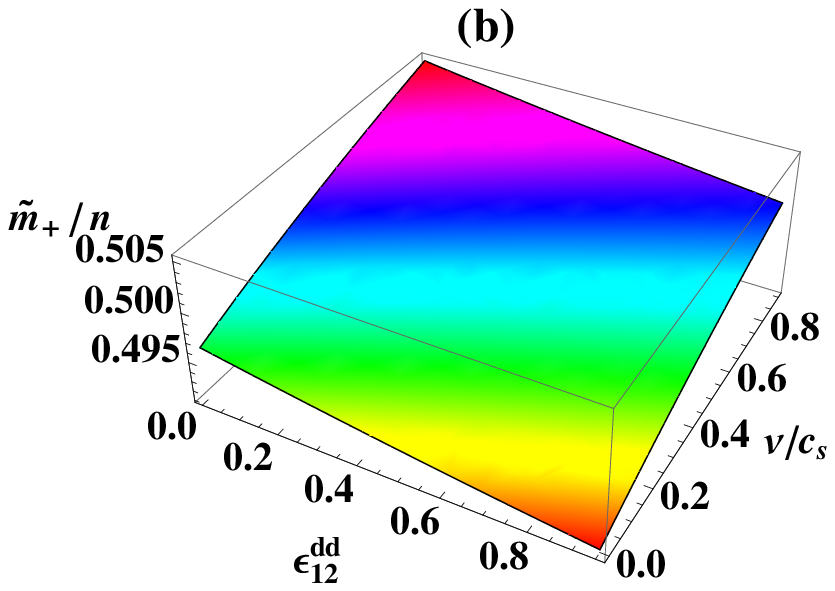}
\includegraphics[scale=0.45]{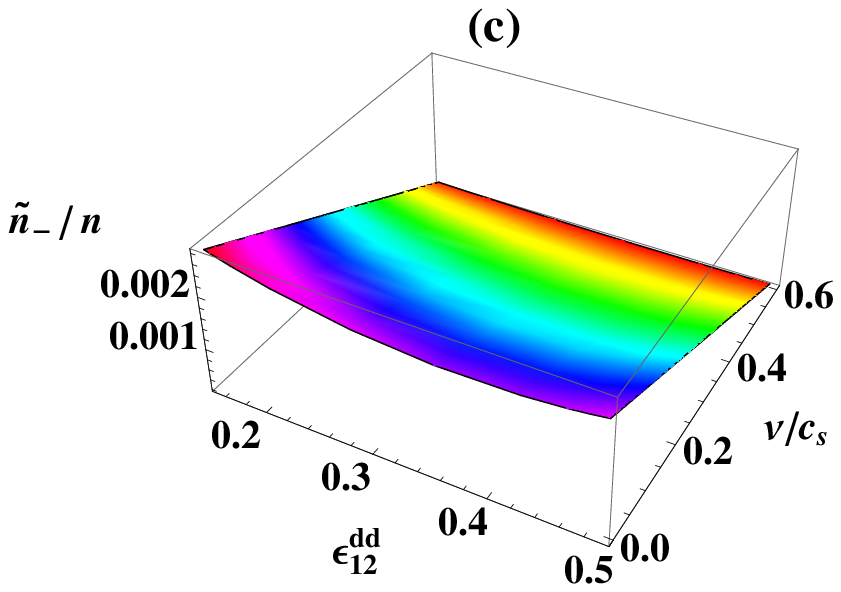}
\includegraphics[scale=0.45]{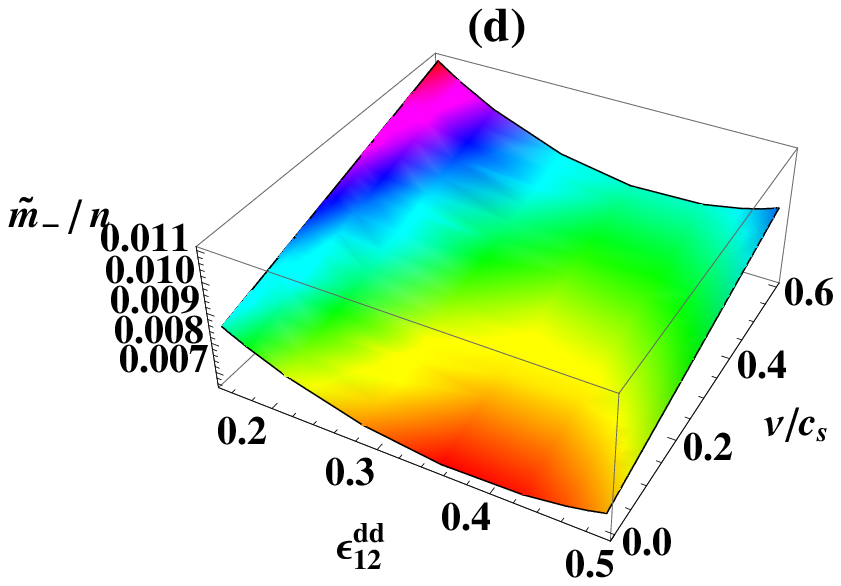}
\caption {(Color online)  (a) Noncondensed density associated with the upper branch as a function of $\epsilon_{12}^{dd}$ and $v/c_s$.
(b) Anomalous density associated with the upper branch as a function of $\epsilon_{12}^{dd}$ and $v/c_s$.
(c) and (d)  the same but for the lower branch.
Parameters are: $a=141 a_0$ \cite{Tang}, $a_{12}=115.1 a_0$, $r_*=131 a_0$ \cite{Pfau1},  and $ n \approx 10^{20} m^{-3}$.}
\label{Dens}
\end{figure}

Figures \ref{Dens}.(a) and (c) show that the noncondensed density in both components is decreasing with the velocity of two fluids leading to 
a large condensed fraction.
We see also that $\tilde n_+$ increases with $\epsilon_{12}^{dd}$ while $\tilde n_-$ first decreases for $\epsilon_{12}^{dd} \lesssim 0.4$ and then starts to lift.
The balance between the DDI and the relative motion drive to the formation of a robust Bose mixture even in the presence of relatively large DDI in contrast to
immovable dipolar systems.
If the velocity of the mixture reaches the Landau critical velocity, then one can expect that the quantum depletion would be significant. 

The situation is different for the anomalous density. Figure \ref{Dens}.(b) depicts that $\tilde m_+$ rises with $v$ and decays with $\epsilon_{12}^{dd}$. 
Whereas,  $\tilde m_-$ changes its charcter with $\epsilon_{12}^{dd}$, first it lowers in the regime $\epsilon_{12}^{dd} \lesssim 0.4$ and then it augments for $\epsilon_{12}^{dd} > 0.4$ 
as is shown in Fig.\ref{Dens}.(d). This unconventional behavior of the noncondensed and the anomalous densities associated with the lower branch can be attributed to the 
interplay of the interspcies DDI and the relative motion of two BECs.
Moreover, we observe that $\tilde n_-$ and $\tilde m_-$ are always smaller than $\tilde n_+$ and $\tilde m_+$ regardless of the values of $\epsilon_{12}^{dd}$  and $v/c_s$
indicating that their effects in moving symmetric binary mixtures is not really important.

\begin{figure}[htb] 
\includegraphics[scale=0.7]{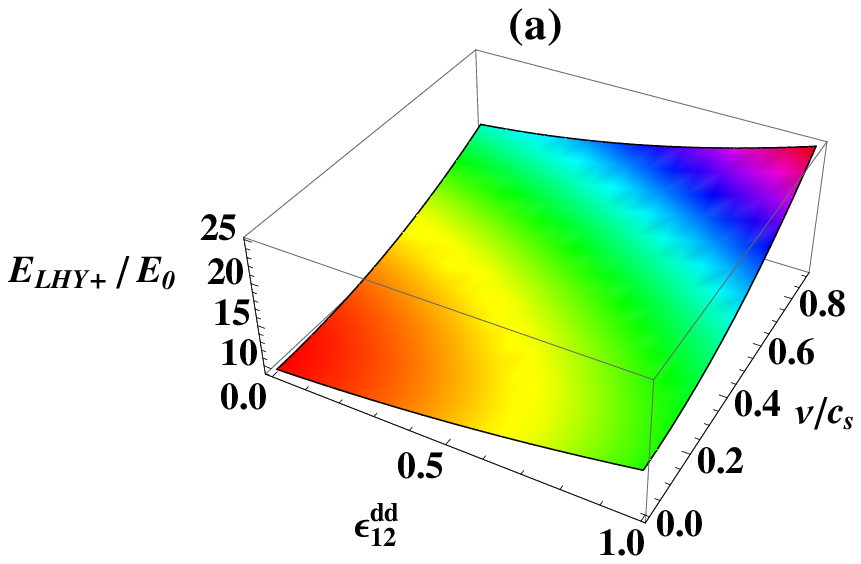}
\includegraphics[scale=0.7]{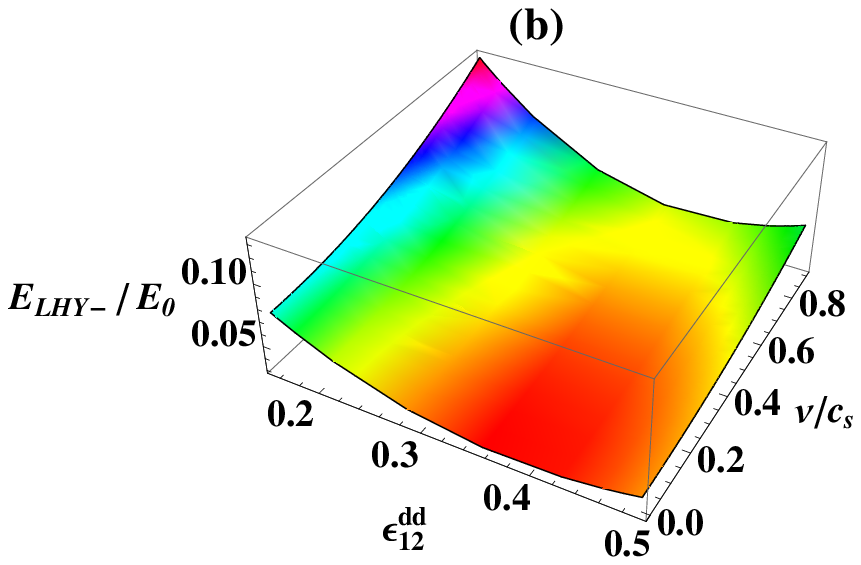}
\caption {(Color online)  
(a) The LHY corrected energy as a function of $\epsilon_{12}^{dd}$ and $v/c_s$.
(b)  the same but for the lower branch.
Parameters are: $a=141 a_0$ \cite{Tang}, $a_{12}=115.1 a_0$, $r_*=131 a_0$ \cite{Pfau1},  and $ n \approx 10^{20} m^{-3}$.}
\label{Egy}
\end{figure}

Figures \ref{Egy}.(c) and (d)  show that the energies in both branches $E_{\text{LHY}\pm}$ increase with $v$.
It is clearly visible that $E_{\text{LHY}+}$ increases with $\epsilon_{12}^{dd}$ while $E_{\text{LHY}-}$ changes its behavior from small to large  $\epsilon_{12}^{dd}$. 
Here again we see that  $E_{\text{LHY}-}$ is negligeable  compared to $E_{\text{LHY}+}$.

Let us now look at how the ground-state energy of each component behaves in the regime of small velocities.
Performing integral (\ref{Energy}) up to second-order in $v$, and using the dimensional regularization \cite{Boudj5} we get
\begin{equation}\label{Energy3}  
E_{\text{LHY}\pm}  \simeq E_0 \, {\cal I}_{\pm 5} (\epsilon^{dd}) +\frac{1}{2} M^*v^2,
\end{equation}
where  $M^*= 15 E_0 {\cal I}_{\pm 5} (\epsilon^{dd})/(8 c_s^2)$, and the dipolar functions ${\cal I}_{\pm 5}(\epsilon_{dd})$ of the ground-state energy
associated with the upper/lower branches in terms of the relative interspecies dipolar interaction strength $\epsilon_{12}^{dd}$ are defined as \cite{Boudj4}: 
\begin{eqnarray} \nonumber 
{\cal I}_{\ell \pm} (\epsilon^{dd})&= \int_0^{\pi}  \sin \theta \left[1+\epsilon^{dd} (3\cos^2\theta-1) \right]^{\ell/2}  \\
& \times \bigg \{  1\pm \frac{ g_{12}\left[1+\epsilon_{12}^{dd} (3\cos^2\theta-1) \right]^{\ell/2}}{g\left[1+\epsilon^{dd} (3\cos^2\theta-1) \right]^{\ell/2}} \bigg \} \, d \theta, \nonumber 
\end{eqnarray}
with $\ell=5$.
For $\epsilon_{12}^{dd}=0$,  the functions ${\cal I}_{\pm 5}(\epsilon^{dd})$ reduces to ${\cal Q}_{5}(\epsilon^{dd})$ describing the DDI for a single BEC.
For $v=0$, one recovers the energy of an immovable dipolar Bose mixture \cite{Boudj5}.
Equation (\ref{Energy3}) shows that the dipolar mixture fluid gains an extra kinetic energy $M^*v^2/2$ due to the relative motion 
leading to increase  the total energy of the mixture.

\section{Second-order correlation function} \label{PCF}

The second-order (pair) correlation function, $g^{(2)} ({\bf r_1},{\bf r_2})=\langle \hat\psi^\dagger({\bf r_1})  \hat\psi^\dagger({\bf r_2})\hat\psi ({\bf r_2})\hat\psi({\bf r_1}) \rangle$,
is an important quantity to characterize the coherence of the self-bound droplet state.
Spliting the bosons field operator $\hat\psi ({\bf r})=\hat {\bar \psi} ({\bf r})+ \Phi ({\bf r})$, where $\hat {\bar \psi} ({\bf r})$ corresponds to quantum 
fluctuations, then using the Wick's theorem, we find for the pair correlation function in each component:
$g_{\pm}^{(2)} ({\bf r_1},{\bf r_2})= n_c^2+2n_c \left[\tilde{n}_{\pm} ({\bf r_1},{\bf r_2})+\tilde{m}_{\pm} ({\bf r_1},{\bf r_2}) \right]$, which
explicitly couples to normal and anomalous correlations. Working in momentum space, we finally get
\begin{equation}\label {2Corr2}
g_{\pm}^{(2)} ({\bf r})=n_c^2+2n_c\int_0^{\infty}  \frac{d {\bf p} }{(2\pi \hbar)^3} \bigg (\frac{ E_{\bf p}'} {\varepsilon_{{\bf p} \pm}}  -1\bigg)  e^{i {\bf p} \cdot {\bf r}/\hbar},
\end{equation}
where ${\bf r}=|{\bf r_1}-{\bf r_2}|$.
Certainly, the absence of $\tilde m$ may affect the long-range behavior of $g^{(2)}(r)$.

\begin{figure}[htb] 
\includegraphics[scale=0.45]{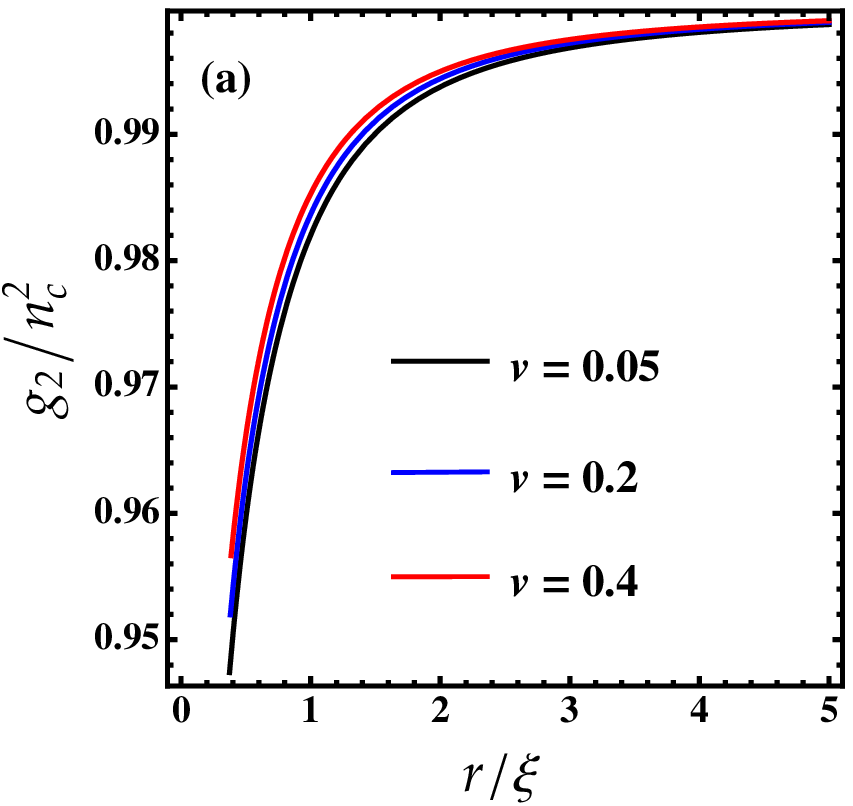}
\includegraphics[scale=0.45]{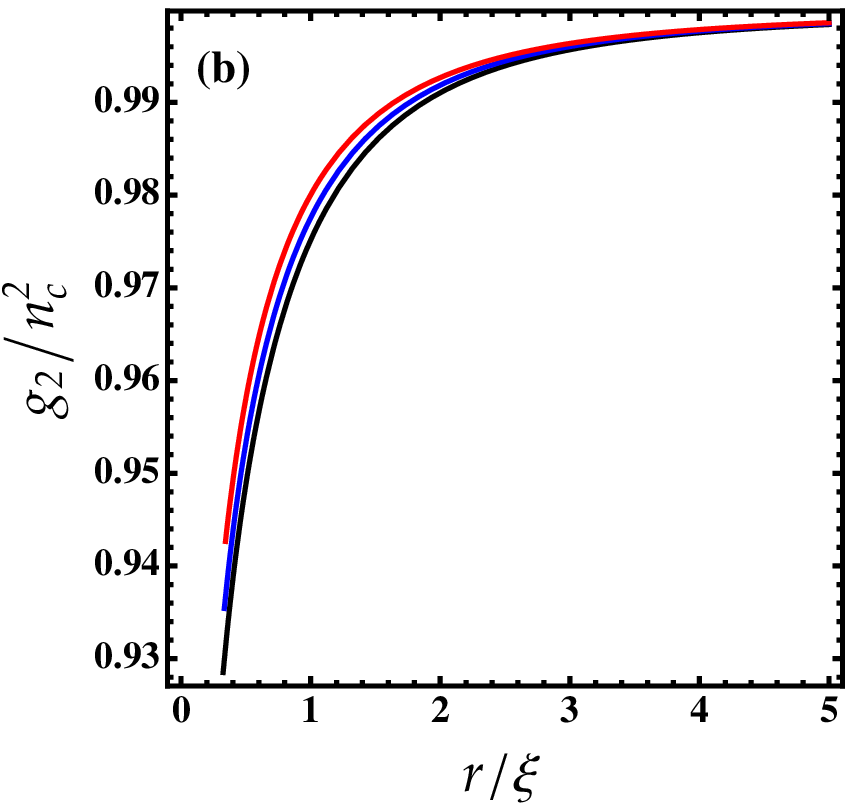}
\includegraphics[scale=0.45]{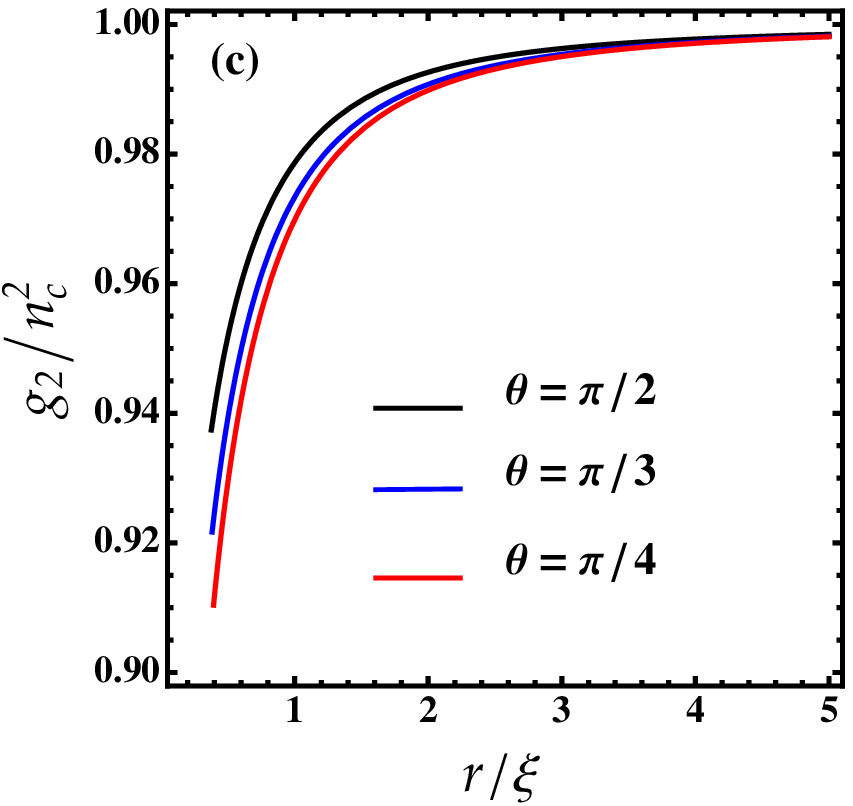}
\includegraphics[scale=0.45]{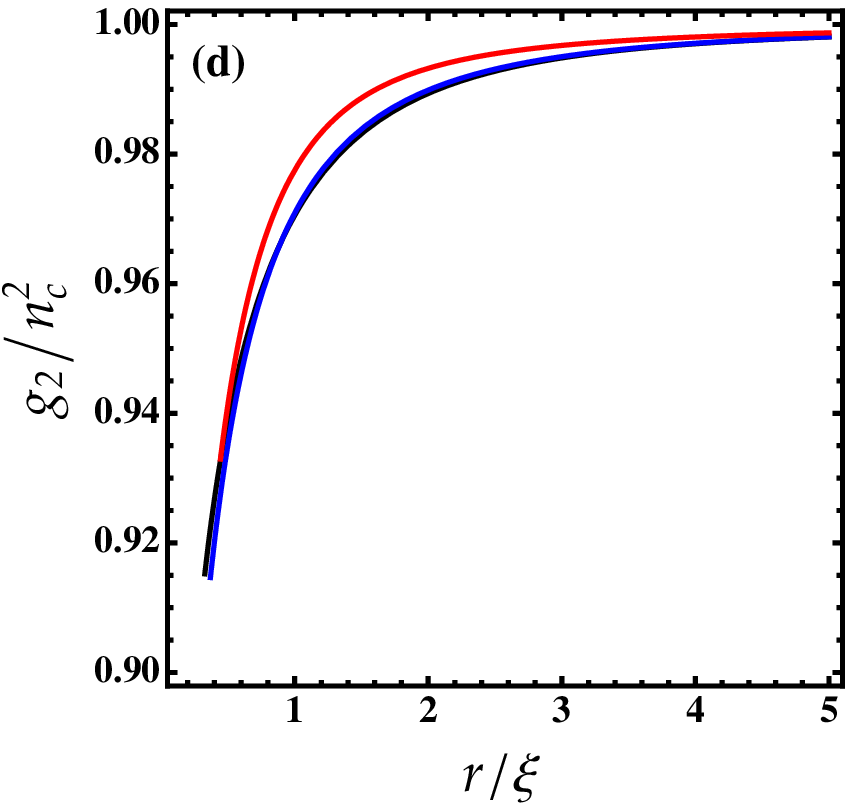}
\caption {(Color online)  (a) Total second-order correlation function $g_{2}(r)/n^2=\sum_{\pm} g_{2\pm}(r)/n^2$ for different values of $v/c_s$ and for $\theta=\pi/2$, 
$\epsilon_{12}^{dd}=0.2$ (a) and $\epsilon_{12}^{dd}=0.8$ (b).
Correlation function $g_{2}(r)/n^2$ for different values of the polarization direction $\theta$ and  for $v/c_s=0.1$, $\epsilon_{12}^{dd}=0.2$ (a) and $\epsilon_{12}^{dd}=0.8$ (b).
Parameters are: $a=141 a_0$, $a_{12}=115.1 a_0$, $r_*=131 a_0$,  and $ n \approx 10^{20} m^{-3}$. }
\label{2cor}
\end{figure}

Figure \ref{2cor} depicts the behavior of the total pair correlation function $g_{2}({\bf r})/n^2=\sum_{\pm} g_{2\pm}({\bf r})/n^2$.
We see that $g_{2}(r)$ vanishes at short distances and then increases monotonically regardless of the values of $v/c_s$ and the polarization direction $\theta$. 
It rises slightly with the velocity of two fluids and remains almost insensitive to the interspecies relative interaction strength $\epsilon_{12}^{dd}$ (see Figs.\ref{2cor} (a) and (b)).
The polarization direction $\theta$ may also lead to increase $g_{2}(r)$ notably for small  $\epsilon_{12}^{dd}$  (see Fig.\ref{2cor} (c)).
Whereas, the situation is inverted for large  interspecies relative interaction strength, $\epsilon_{12}^{dd}=0.8$, (see Fig.\ref{2cor} (d)).
Experimentally, the pair correlation function can be measured using either a Bragg diffraction interferometer \cite{Cac} or four-wave mixing
of the collision of two BECs \cite{Perin}.

\section{Conclusion} \label{Conc}

We studied the behavior of moving  uniform dipolar Bose-Bose mixtures using the full HFB theory.
This latter  is valid only in the dilute regime and for small velocity.
We calculated experimentally relevant quantities such as the depletion and the ground-state energy for both single and binary BECs. 
We found that the relative motion corrections to the energy provide a term quadratic in $v$ leaing to rise the total energy of the fluid.
The pair correlation function has been also evaluated in different regimes.
Our analysis revealed that the intriguing interplay of the relative motion of two components and DDI may affect the stability and the coherence of the mixture. 
The results of the present work pave the way for the simulation of complex many-body systems such as quantum self-bound droplet in moving Bose mixtures.
In the case of a dipolar Bose-Bose mixture with spin-orbit coupling (SOC), one can expect that the mobility of the system becomes a nontrivial 
since the SOC terms break the Galilean invariance of the generalized nonlocal coupled GP equations \cite{Mard,Jiang}. 
It may be also interesting to explore the impact of SOC in dipolar Bose mixtures as future extension of the present work.


\subsection*{Data availability statement}
The data generated and/or analyzed during the current study are not publicly available for legal/ethical reasons
but are available from the corresponding author on reasonable request.

\subsection*{Author contribution statement}
All authors discussed the results and made critical contributions to the work. AB contributed to the writing of the manuscript.

\end{document}